# Application of effective medium theory to estimate gas permeability in tight-gas sandstones

Behzad Ghanbarian[1*], Carlos Torres-Verdín[2], Larry W. Lake[2] and Michael Marder[3]

[1] Department of Geology, Kansas State University, Manhattan 66506 KS, USA

[2] Department of Petroleum and Geosystems Engineering, University of Texas at Austin, Austin 78712 TX, USA

[3] Department of Physics, University of Texas at Austin, Austin 78712 TX, USA

[*] Corresponding author's email address: ghanbarian@ksu.edu

**Abstract**

Upscaling Klinkenberg-corrected gas permeability, $k$, in unconventional tight sandstones has numerous practical applications, particularly in gas exploration and production. In this study, we adapt the effective-medium approximation (EMA) model of *Doyen* – proposed first to estimate bulk electrical conductivity, $\sigma_b$, and permeability in sandstones from rock images – to scale up $\sigma_b$ and $k$ in tight-gas sandstones from pore to core. For this purpose, we calculate two characteristic pore sizes: an effective hydraulic and an effective electrical pore size from pore-throat size distributions – determined from mercury intrusion capillary pressure (MICP) curves – and pore-throat connectivity. The latter is estimated from critical volume fraction (or percolation threshold) for macroscopic flow. Electrical conductivity and permeability are then scaled up from the two characteristic pore sizes, tortuosity, and porosity by assuming two different pore geometries: cylindrical and slit-shaped. Comparison of results obtained for eighteen tight-gas sandstones indicates that the EMA estimates $\sigma_b$ and $k$ more accurately when pores are assumed to be cylindrical. We also estimate $k$ from the pore-throat size distributions and the measured electrical conductivity using the EMA and critical path analysis (CPA), another upscaling technique borrowed from statistical physics. Theoretically, the former is valid in relatively heterogeneous porous media with narrow pore-throat size distribution, while the latter is valid in heterogeneous media with broad pore-throat size distribution. Results show that the EMA estimates $k$ more accurately than CPA and arrives within a factor of two of the measurements on average.

## 1. Introduction

Upscaling gas permeability, $k$, in tight porous sandstones has numerous practical applications in reservoir engineering. Various theoretical models have been proposed to scale up $k$ from pore throat properties, employing size distribution, tortuosity, porosity, etc. (Behrang et al., 2016; Peng et al., 2018; Song et al., 2016). One simple model based on a bundle of tortuous capillary tubes of the same radius $R$ and porosity $\phi$ is given by (Larry et al., 2014)

$$k = \frac{R^2 \phi}{8\tau}. \tag{1}$$

Here tortuosity $\tau = \left(\frac{L_t}{L_s}\right)^2$, where $L_t$ is the tortuous capillary tube length and $L_s$ is the sample length.

In Eq. (1) by replacing $R$ with $2R_h = 2\phi/[S_a(1-\phi)]$ where $R_h$ is hydraulic radius and specific surface area $S_a = 6/\bar{d}_g$, one may derive the Kozeny-Carman model (Carman, 1937; Kozeny, 1927) as

$$k = \frac{\phi^3}{2\tau S_a^2(1-\phi)^2} = \frac{\bar{d}_g^2 \phi^3}{72\tau(1-\phi)^2}. \tag{2}$$

Here $\bar{d}_g$ is a representative grain diameter that may be approximated by the harmonic or geometric mean (Koltermann and Gorelick, 1995). However, as is well known, Eq. (2) only provides accurate permeability estimations in homogeneous porous media with narrow particle/pore size distribution (Bryant et al., 1993).

Generalized models based on a bundle of capillary tubes concept that include the Kozeny-Carman equation as a special case were later developed by incorporating more complex properties of the pore space morphology (see e.g., Behrang and Kantzas (2017); Cai et al. (2015); Chen and Yao (2017)). However, idealizing a porous medium as a bundle of straight/tortuous capillary tubes even with various pore shapes (see e.g., Cai et al., 2014) neglects pore connectivity, which is important for fluid transport through rocks. More complex approaches were also invoked to model gas permeability based on first- and higher-order slip equation. For example, Beskok and Karniadakis (1999) accounted for slip flow and adapted Hagen-Poiseuille's law for gas flow in a cylindrical tube. Such a method was extensively used to model Klinkenberg-corrected gas permeability in unconventional reservoir rocks (see e.g., Civan, 2010; Nazari Moghaddam and Jamiolahmady, 2016). Javadpour (2009) and Darabi et al. (2012), however, accounted for first-order slip flow and Knudsen diffusion mechanisms and proposed another model for gas permeability.

Techniques borrowed from statistical physics have been used to upscale permeability in porous media. For example, Katz and Thompson (Katz and Thompson, 1986; Katz and Thompson, 1987) applied critical path analysis to estimate the permeability of rocks. In their model, a critical pore size is the key factor controlling permeability. Although critical path analysis was proposed to upscale electrical conductivity and permeability in porous media with broad conductance distribution, Shah and Yortsos (1996) stated that, "The basic argument underlying this theory [critical path analysis] is that because of the large exponent in the pore conductance-pore [throat] radius relationship, $g \sim r^4$, natural porous media, even though moderately disordered in pore [throat] size, possess a wide conductance distribution." Concepts from critical path analysis were successfully used to upscale single-phase permeability in different types of rocks and soils (Arns et al., 2005; Ghanbarian et al., 2017, 2016b; Hunt et al., 2014; Katz and Thompson, 1986; Katz and Thompson, 1987).

In addition to critical path analysis, the effective-medium approximation, an old approximate analytical upscaling technique, was also applied to upscale fluid flow and transport in porous rocks (see e.g., Sahimi (2011) and references therein). Koplik et al. (1984) assumed that pore throats were elliptically cylindrical and determined hydraulic and electrical conductances, their distributions, and the average coordination number ($Z$; the average number of pore throats connected to the same pore body) from thin sections. Using the effective-medium approximation, Koplik et al. (1984) overestimated permeability by a factor of 10, while underestimating formation factor ($F = \sigma_f/\sigma_b$ in which $\sigma_b$ and $\sigma_f$ are bulk and fluid electrical conductivities, respectively) by a factor of 2 for a Massilon sandstone with porosity $\phi = 0.22$ and average pore coordination number $Z = 3.5$.

Following Koplik et al. (1984), Doyen (1988) estimated the single-phase permeability of seven Fontainebleau sandstones from thin sections and image analysis. An effective pore-throat radius was determined from pore-throat and pore-body size distributions. Doyen (1988) performed both permeability and electrical conductivity estimations within a factor of three of the measurements. David et al. (1990) assessed the EMA reliability in estimation of effective conductance in two-dimensional regular networks (e.g., hexagonal, square, and triangular) with various pore conductance distributions. Those authors reported that the effective conductance estimated by the EMA was in good agreement with numerical simulations of flow through media with quasi-uniform distributions, while large discrepancies appeared for exponential distributions. Adler and Berkowitz (2000) evaluated the accuracy of the EMA in the estimation of electrical conductivity in two- and three-dimensional lattices with local conductances following a lognormal distribution of various standard deviations. Adler and Berkowitz (2000) concluded that, "… the analytical expressions [the effective-medium approximations] provide good agreement to the simulations in 2D systems, but are in significant error in 3D systems when the standard deviation [$s_{ge}$] of the local conductivities is large." They indicated that EMA conductivity estimations in three dimensions were very accurate under fully saturated conditions as long as the standard deviation of the lognormal electrical conductance distribution was less than 1 ($s_{ge} < 1$).

To scale up single-phase gas permeability from pore to core (e.g. via Eq. 1), one requires knowledge of porosity, tortuosity, shape factor and a characteristic length scale. Although Eq. (1) was proposed for a medium composed of tortuous capillary tubes of uniform radius, $R$, it has been widely used to estimate permeability in homogeneous and heterogeneous porous rocks by replacing $R$ with $\bar{R}$, a representative pore radius in the medium. In the literature, different quantities, such as effective pore-throat radius, $r_e$ (David et al., 1990; Doyen, 1988), and median pore-throat radius, $r_{50}$ (Weller et al., 2016), have been used to estimate $\bar{R}$ and accordingly $k$ via Eq. (1).

The effective-medium approximation has been previously used to estimate $\bar{R}$, (or the effective hydraulic pore radius $r_{he}$) from the pore-throat size distribution and the average pore coordination number $Z$, both determined from *two-dimensional* images of sandstones and carbonate rocks (see e.g., Doyen (1988); Koplik et al. (1984)). However, precise calculation of $Z$ requires either two- or three-dimensional rock images or nitrogen sorption measurements (Seaton, 1991). We accordingly approximate $Z$ from the critical saturation, $S_c$, corresponding to the inflection point on the capillary pressure curve, routinely measured through petrophysical evaluation of porous rocks. Although $Z$ approximation from $S_c$ has been successfully tested on the estimation of relative permeability in soils (Ghanbarian et al., 2016a) and pore-pressure dependent gas permeability in shales (Ghanbarian and Javadpour, 2017), to the best of the authors' knowledge, it has never been used to estimate the effective pore-throat size and the Klinkenberg-corrected gas permeability from mercury intrusion porosimetry *in tight-gas sandstones*. Nor has it been applied to estimate gas permeability from electrical conductivity measurements. Furthermore, its accuracy has not been compared to that of critical path analysis *experimentally in porous media*. Therefore, the main objectives of this study are: (1) to evaluate the EMA's reliability and accuracy to estimate effective hydraulic and electrical pore sizes and the Klinkenberg-corrected gas permeability $k$ from pore-throat size distribution, tortuosity, and porosity, (2) to estimate $k$ from pore-throat size distribution and electrical conductivity measurements, and (3) to compare EMA results to those obtained from critical path analysis in tight-gas sandstones.

## 2. Theory

### 2.1. Hydraulic and electrical conductances of a single pore

To scale up gas permeability and electrical conductivity in a low permeability (tight) porous rock, one first requires describing hydraulic and electrical flow at the pore scale for a single pore. The hydraulic ($g_h$) and electrical ($g_e$) conductances of a cylindrical pore of radius $r$ with a constant cross-sectional area and length $l$ filled with a fluid of viscosity $\mu$ and electrical conductivity of $\sigma_f$ are (Friedman and Seaton, 1998)

$$g_h = \frac{\pi r^4}{8\mu l} \tag{3}$$

and

$$g_e = \sigma_f \frac{\pi r^2}{l} \tag{4}$$

For a slit-shaped pore of width $w$ much narrower than its breadth $b$ and length $l$, hydraulic and electrical conductances are (Friedman and Seaton, 1998)

$$g_h = \frac{bw^3}{12\mu l} \tag{5}$$

and

$$g_e = \sigma_f \frac{bw}{l} . \tag{6}$$

Note that the numerical prefactors 8, in Eq. (3), and 12, in Eq. (5), are pore shape factors.

**2.2. Scaling up from pore to core via the effective-medium approximation**

Within the EMA framework, a heterogeneous and uncorrelated pore network is replaced by an ordered one composed of pores all of the same size, while its permeability is still the same as that of the original disordered network.

Figure 1 presents a 3D schematic pore network with pore bodies and throats of various sizes replaced by an ordered network (lattice) with a single effective pore size and conductance. Following Kirkpatrick (1973), the effective pore conductance is determined by setting the spatially averaged perturbations, taken with respect to the conductance distribution $f(g)$ in the disordered network, to be zero as follows

$$\int_{g_{min}}^{g_{max}} \frac{(g_e-g)}{g+(Z/2-1)g_e} f(g)dg = 0 \qquad (7)$$

where $g_{min}$ and $g_{max}$ are the minimum and maximum conductances in the network, $g_e$ is the effective conductance controlling flow, and $Z$ is the average pore coordination number, (the average number of pore throats meeting at a pore body). Hereafter, we refer to pore-throat and pore-body radii as $r$ and $r_b$, respectively, in cylindrical pores and as $w$ and $w_b$ in slit-shaped pores.

Three-dimensional images provide detailed pore space geometrical and morphological properties, particularly compared to other methods like mercury intrusion porosimetry. Although different algorithms e.g., Lindquist et al. (1996), Silin et al. (2003) and Øren and Bakke (2003) have been developed to quantify pore space geometrical properties and topological characteristics in porous media, analysis of three-dimensional images and identification of pore radius, volume, and length are still ambiguous. Note that widely-applied and well-known models characterizing pore space from rock images, such as the maximal ball algorithm (Dong and Blunt, 2009; Silin and Patzek, 2006) and 3DMA-Rock (Lindquist et al., 2005; Lindquist et al., 2000) still differentiate pore bodies from pore throats.

Within the effective-medium approximation framework, the bond percolation threshold ($p_c$) of a pore network is $2/Z$. Although $p_c = 2/Z$ is accurate in two dimensions, it has been well documented in the literature (see, e.g., (Kirkpatrick, 1973)) that the EMA, Eq. (7), generally overestimates $p_c$ in 3D networks ($p_c < 2/Z$). In addition, accurate estimation of $Z$ requires either 2D or 3D rock images (Doyen, 1988) or a nitrogen adsorption curve (Seaton, 1991). Because neither the sorption isotherms nor the images for the samples studied here are available, we use another approximation and replace $2/Z$ in Eq. (7) with $p_c$, i.e. $(Z/2 - 1) \approx (1 - p_c)/p_c$. As suggested by Katz and Thompson (Katz and Thompson, 1986; Katz and Thompson, 1987), the saturation at the inflection point ($S_c$) on the mercury intrusion capillary pressure (MICP) curve represents the critical volume fraction (the minimum saturation required to form sample-spanning cluster) for percolation. The bond percolation threshold ($p_c$) representing the critical number fraction is equivalent to the critical mercury saturation ($S_c$) representing the critical volume fraction. Following Ghanbarian et al. (2016a), we thus correspondingly replace $(1 - p_c)/p_c$ by $(1 - S_c)/S_c$ and rewrite Eq. (7) to obtain

$$\int_{g_{min}}^{g_{max}} \frac{(g_e - g)}{g + [(1-S_c)/S_c]g_e} f(g) dg = 0 \qquad (8)$$

Given that $g_h \propto r^4$, $g_e \propto r^2$, and $f(g)dg = f(r)dr$, we accordingly substitute the corresponding conductance from Eqs. (3) and (4) into Eq. (8) to determine the effective pore radius for, respectively, hydraulic and electrical flow from the pore-throat size distribution $f(r)$ and the critical saturation $S_c$. Similarly, for slit-shaped pores Eqs. (5) and

(6) are substituted in Eq. (8) to calculate $w_{he}$ and $w_{ee}$, the effective hydraulic and electrical pore widths, respectively.

In accordance with Doyen (1988), the macroscopically upscaled electrical conductivity $\sigma_b$ and permeability $k$ are given by

$$\frac{\sigma_b}{\sigma_f} = \frac{\phi}{\tau_e} \frac{r_{ee}^2}{\langle r_b^2 \rangle} \tag{9}$$

and

$$k = \frac{\phi}{8\tau_h} \frac{r_{he}^4}{\langle r_b^2 \rangle} \tag{10}$$

where $\tau_e$ and $\tau_h$ are, respectively, the electrical and hydraulic tortuosity factors. Therefore, one may estimate the electrical conductivity and permeability from the effective pore radii ($r_{he}$ and $r_{ee}$), the average squared pore body radius ($\langle r_b^2 \rangle$), tortuosity, and porosity. In Eqs. (9) and (10), the value of $\langle r_b^2 \rangle$ is determined as the average of $r_b^2$ over the entire pore body-size distribution. In this study, the pore body-size distribution is not available. We thus *approximate* $\langle r_b^2 \rangle$ with $\langle r^2 \rangle$, the average squared pore-throat radius (David et al., 1990; Doyen, 1988). According to the results of Zhao et al. (2015) and Xi et al. (2016), replacing $\langle r_b^2 \rangle$ with $\langle r^2 \rangle$, however, might be a *rough* approximation in tight sandstones.

Dividing Eq. (10) by (9) gives

$$k = \frac{\tau_e}{8\tau_h} \frac{\sigma_b}{\sigma_f} \frac{r_{he}^4}{r_{ee}^2} \tag{11}$$

Ghanbarian et al. (2013a) demonstrated that electrical tortuosity $\tau_e$ is typically smaller than hydraulic tortuosity $\tau_h$. However, if the pore-throat size distribution is narrow enough, one may approximately set $\tau_e \approx \tau_h$ (Ghanbarian et al., 2013a) and, accordingly, Eq. (11) becomes

$$k = \frac{1}{8}\frac{\sigma_b}{\sigma_f}\frac{r_{he}^4}{r_{ee}^2} \qquad (12)$$

Equation (12) estimates $k$ from the measured electrical conductivity and the effective hydraulic and electrical pore-throat radii (i.e., $r_{he}$ and $r_{ee}$).

Likewise, for slit-shaped pores Eqs. (9), (10) and (12) become

$$\frac{\sigma_b}{\sigma_f} = \frac{\phi}{\tau_e}\frac{w_{ee}}{\langle w_b \rangle} \qquad (13)$$

$$k = \frac{\phi}{12\tau_h}\frac{w_{he}^3}{\langle w_b \rangle} \qquad (14)$$

and

$$k = \frac{1}{12}\frac{\sigma_b}{\sigma_f}\frac{w_{he}^3}{w_{ee}} \qquad (15)$$

Recall that $w_{he}$ and $w_{ee}$ are, respectively, the effective hydraulic and electrical pore widths, and $w_b$ is the pore body width.

In what follows, we apply Eqs. (9), (10) and (12) as well as Eqs. (13), (14) and (15) in order to estimate electrical conductivity and gas permeability in eighteen tight-gas sandstone samples. We compare the accuracy of the effective-medium approximation to that of critical path analysis in the $k$ estimation from the pore-throat size distribution and electrical conductivity measurements.

## 3. Experimental data

The eighteen tight-gas sandstones used in this study were cut from whole cores retrieved from a tight-gas sandstone formation located in East Texas. In all samples, gas permeability was measured using the transient pulse technique at a net confining stress of 2500 psi and corrected by extrapolating to infinite pressure using the Klinkenberg (1941) method. The effect of pore and confining pressures on fluid permeability has been extensively addressed in porous rocks (see e.g., Brace and Walsh (1968); Chen et al. (2015); Cui et al. (2018); Fink et al. (2017); Jasinge et al. (2011); Klinkenberg (1941); Walsh (1981); Wu et al. (2017)). However, the influence of pore pressure on gas permeability in our samples is small because all permeability measurements were Klinkenberg-corrected. Total porosity of each sample was determined via helium expansion. The capillary pressure curve, measured using mercury intrusion pressures ranged between 1.5 to 60000 psi, was used to derive the pore-throat size distribution i.e., $f(r)$ or $f(w)$ for each sample. For this purpose, capillary pressure $P_c$ was converted to pore-throat diameter ($d = 2r$) and/or width ($w$) using respectively (Bullard and Garboczi, 2009)

$$P_c = P_{nw} - P_w = \frac{4\gamma \cos\theta}{d}, \qquad (16a)$$

$$P_c = P_{nw} - P_w = \frac{2\gamma \cos\theta}{w}, \qquad (16b)$$

in which $P_{nw}$ and $P_w$ are the pressures of the nonwetting and wetting phases, respectively, $\gamma$ is the air/mercury interfacial tension (485 dyn/cm), and $\theta$ is the contact angle (140° for mercury). Equation (16) necessarily means that pore radius is equivalent to pore width ($r = w$) and thus $f(r) = f(w)$. To determine the pore-throat size distribution, $dS_{Hg}/d\ln(P_c)$ was calculated from the MICP and plotted versus pore-throat size, as shown in Fig. 2. In this study, MICP and permeability were measured on the same sample. The core samples measured 1 inch in diameter and 1.5 inches in length. It is well documented in the literature that petrophysical properties such as permeability, porosity, and MICP (or pore-throat size distribution) are scale-dependent measurements. For example, both Tinni et al. (2012) and Fink et al. (2017) experimentally showed that as sample volume increases, permeability increases as well. Results of Chen et al. (2015) and Han et al.

(2016) also indicated that the pore-throat size distribution could vary with a change in measurement scale.

Following Katz and Thompson (Katz and Thompson, 1986; Katz and Thompson, 1987), the critical saturation $S_c$ for gas flow was estimated from the inflection point on the MICP curve. We determined the inflection point in MATLAB by fitting a spline to the measured data and computing the second derivative numerically. However, the method failed to correctly find the inflection point because of local scatter in MICP data in samples 9, 10, 13, 16, and 18. We instead fit the van Genuchten capillary pressure curve model (van Genuchten, 1980) and then calculated the inflection point corresponding to the critical saturation $S_c$, critical pore radius $r_c$, and critical pore width $w_c$ using the approach proposed by Dexter (2004) and implemented by Ghanbarian et al. (2016b). Using $f(r)$, $f(w)$, the critical saturation $S_c$, and Eq. (8) we calculated the effective hydraulic pore sizes i.e., $r_{he}$ and $w_{he}$ as well as the effective electrical i.e., $r_{ee}$ and $w_{ee}$.

Surface conduction can contribute to electrical conductivity in addition to bulk conduction in clay-rich and clay-free porous rocks (see e.g., Revil and Glover (1998); Revil et al. (2014)). In this study, electrical conductivity was measured using a synthetic and *highly saline* brine of resistivity 0.0403 ($\Omega$-m) at 25°C using the two-terminal method and 1 KHz frequency under fully saturated conditions. Such a low resistivity corresponds to high brine conductivity $\sigma_f = 24.8$ (S m$^{-1}$), high bulk electrical conduction, and, therefore, negligible surface conduction ($\sigma_b \gg \sigma_s$), in accord with results given in Fig. 3 of Revil et al. (2014). In addition, measurements of electrical conductivity, particularly at small water saturations (not shown), indicated that surface conduction was negligible in these tight-gas sandstones. This behavior is also confirmed through accurate permeability estimations from the measured electrical conductivity, as shown in the Results and Discussion section.

To estimate permeability and electrical conductivity via Eqs. (9), (10), (13) and (14), one needs to estimate electrical $\tau_e$ and hydraulic $\tau_h$ tortuosity values. Analysis of energy or Joule dissipation (Bernabé and Revil, 1995; Revil and Cathles, 1999) has shown that hydraulic or electrical tortuosity is a dynamic quantity associated with the local normalized gradient of the fluid pressure or electrical potential. However, three-dimensional images of the tight-gas sandstones studied here are unavailable to provide an accurate determination of the corresponding tortuosity for hydraulic and electrical flow. Therefore, as a first-order approximation we invoke the geometrical model of Ghanbarian et al. (2013b) and approximate $\tau_e$ and $\tau_h$ with the geometrical tortuosity $\tau_g$

$$\tau_g = \left(\frac{L_e}{L_s}\right)^2 = \left[\frac{\phi - \phi S_c + (C/L_s)^{\frac{1}{\nu}}}{1 - \phi S_c}\right]^{2(\nu - \nu D_{opt})} \tag{17}$$

where $\nu = 0.88$ is the correlation length exponent from percolation theory, $C$ is the typical pore-throat length, $L_e$ is the effective geometrical flow length, and $L_s$ is the sample length. In Eq. (17), $D_{opt}$ is the optimal path fractal dimension whose universal value in 3D systems is 1.43. This means that its value is only a function of system's dimensionality and does not vary from one rock sample to another. Note that the optimal path is the most energetically favorable path through a system. Since $L_s \gg C$ in experiments, following Ghanbarian et al. (2013b) one may set $C/L_s = 0$. By comparing to numerical simulations and experimental measurements, Ghanbarian et al. (2013b) showed that their geometrical tortuosity model, Eq. (17), approximated hydraulic and electrical tortuosity values accurately in porous media with narrow pore/particle size distributions (see their Figs. 3-6). Table 1 summarizes the geometrical tortuosity $\tau_g$ determined for each rock sample. Other salient properties of each sample as well as the calculated parameters to estimate $\sigma_b/\sigma_f$ and $k$ are included in Table 1.

To evaluate the accuracy of the models in the estimation of $\sigma_b/\sigma_f$ and $k$, the root mean square log-transformed error (RMSLE) and mean log-transformed error (MLE) parameters were determined as follows

$$RMSLE = \sqrt{\frac{1}{N}\sum_{i=1}^{N}[\log_{10}(x) - \log_{10}(y)]^2} \tag{18}$$

$$MLE = \frac{1}{N}\sum_{i=1}^{N}[\log_{10}(x) - \log_{10}(y)] \tag{19}$$

where $N$ is the number of values, and $x$ and $y$ are the calculated (estimated) and measured quantities, respectively. Note that MLE > 0 indicates that the model generally overestimates, while MLE < 0 denotes model underestimation.

## 4. Results and Discussion

Table 1 summarizes the calculated effective hydraulic and electrical pore-throat sizes under two different pore geometries (i.e., cylindrical and slit-shaped). We found that except for one of the samples the effective hydraulic radius ($r_{he}$) was greater than the effective electrical radius ($r_{ee}$). In sample 4, $r_{ee} = 0.213$ is slightly greater than $r_{he} = 0.211$, whereas $w_{ee} = 0.211$ is less than $w_{he} = 0.212$. Although Doyen (1988) also reported $r_{he} < r_{ee}$ for two samples in his study, the difference between $r_{ee} = 0.213$ and $r_{he} = 0.211$ in sample 4 is not significant and the two radii can be considered identical with two-decimal-place accuracy. One should typically expect $r_{he} > r_{ee}$ because hydraulic conductance $g_h$ is proportional to pore size (i.e., $r$ or $w$) to the power 4 or 3 for cylindrical and slit-shaped pores, respectively (see Eqs. 3 and 5), while the power for the electrical conductance $g_e$ is 2 or 1 (see Eqs. 4 and 6). This necessarily means that permeability depends on pore space morphological properties more strongly than does electrical conductivity.

Figure 2 shows the pore-throat size distribution, the interpolated curve via the spline method (shown in blue), and the effective hydraulic and electrical pore-throat widths for four samples e.g., 3, 8, 13, and 18. This figure indicates how close the effective hydraulic and electrical widths are to each other and particularly to the peak on the pore-throat size distribution. The peak corresponds to the critical pore-throat width as suggested by Katz and Thompson (Katz and Thompson, 1986; Katz and Thompson, 1987).

We also found a strong correlation between $r_{he}$ and $r_{ee}$ (i.e., $r_{he} = 1.075 r_{ee} - 0.005$, $R^2 = 0.99$) as well as $w_{he}$ and $w_{ee}$ (i.e., $w_{he} = 1.095 w_{ee} - 0.004$, $R^2 = 0.99$). Interestingly, by analyzing the results of Doyen (1988), obtained from images and given in his Table 1, and the results of David et al. (1990), obtained from 2D lattices and presented in their Table 2 and Figs. 3 and 4, we also found that $r_{he}$ was highly correlated to $r_{ee}$ according to $r_{he} = 1.194 r_{ee} - 1.787$ with $R^2 = 0.99$ and $r_{he} = 1.07 r_{ee} - 0.353$ with $R^2 = 0.98$, respectively. In our study, $r_{he}$ and $r_{ee}$ (in μm) were determined from volume-based pore-throat size distribution (e.g., the cumulative volume of mercury injected), while in Doyen (1988) $r_{he}$ and $r_{ee}$ (in μm) were calculated from area-based pore-throat size distribution derived from two-dimensional images. Nonetheless, plotting all $r_{he}$ values versus $r_{ee}$ values from Doyen (1988), David et al. (1990), and this study resulted into the relationship: $r_{he} = 1.06 r_{ee} - 0.097$ with $R^2 = 0.99$ (results not shown). This result indicates that $r_{he}$ is a linear function of $r_{ee}$ to high accuracy. High correlation between $r_{he}$ and $r_{ee}$ as well as $w_{he}$ and $w_{ee}$ might be due to the fact that both parameters are determined from Eq. (8) with the same input parameters, such as pore-throat size distribution, critical saturation, and minimum and maximum pore conductances except the exponent in the conductance-size relationship (see exponents in Eqs. (3)-(6)). Depending on transport mechanism (hydraulic or electrical) and pore geometry (cylindrical or slit-shaped) the exponent could be 1, 2, 3 or 4. As can be observed in Table 1, the values of $r_{ee}$, $r_{he}$, $w_{ee}$, and $w_{he}$ are not greatly different for each rock sample. This behavior indicates that the impact of the exponent in the pore conductance-size relationship on the effective hydraulic and/or electrical pore size estimation is not significant. However, its effect on the estimation of electrical conductivity $\sigma_b/\sigma_f$ and permeability $k$ may be considerable (compare Fig. 3a to Fig. 4a and Fig. 3b to Fig. 4b). This is because the exponent of the effective pore size in Eq. 9 is two (Fig. 3a), whereas its value in Eq. 13 is one (Fig. 4a). Likewise, the exponent of the effective pore-throat size in Eq. 10 is four (Fig. 3b), while its value in Eq. 14 is three (Fig. 4b).

In what follows we first present the results of electrical conductivity and permeability estimations using the effective-medium approximation models i.e., Eqs. (9), (10), and (12)-(15) for two different pore geometries: cylindrical and slit-shaped. We then compare the effective-medium approximation with the critical path analysis method in the estimation of gas permeability from the pore-throat size distribution and electrical conductivity measurements in tight-gas sandstones.

**4.1. Effective-medium approximation**

Results of the effective-medium approximation in the estimation of electrical conductivity $\sigma_b/\sigma_f$ and permeability $k$ are presented in Figs. 3 and 4 assuming pores are cylindrical and slit-shaped, respectively. Figs. 3a, 3b, 4a and 4b show EMA estimations from the pore-throat size distribution, porosity and tortuosity, while Figs. 3c and 4c refers to EMA estimates from the pore-throat size distribution and the measured electrical conductivity. Figure 3a shows that Eq. (9) generally overestimated the electrical conductivity with RMSLE = 0.53 and MLE = 0.48. This is in contrast with the results of Doyen (1988) who showed that the EMA generally underestimated $\sigma_b/\sigma_f$ from pore-throat size distribution derived from two-dimensional images (see his Table 1). Generally speaking, derivation of a three-dimensional quantity from two-dimensional images is inherently difficult because: (1) the interpretation of pore bodies and pore throats is ambiguous, and (2) such images do not capture pore connectivity accurately in all directions, particularly in anisotropic porous rocks. We should point out that Doyen (1988) approximated tortuosity by 3 in all sandstone samples with a wide range of porosity from 5 to 22%, although it is well documented in the literature that tortuosity varies with porosity. Generally speaking, the higher the porosity the less the tortuosity (recall that $\tau \geq 1$).

Equation (9) is an integrated function of several factors and precise estimation of $\sigma_b/\sigma_f$ by Eq. (9) requires accurate determination of each factor i.e., tortuosity, $\langle r_b^2 \rangle$, and effective electrical radius $r_{ee}$. Given that the electrical tortuosity $\tau_e$ was probably underestimated by the geometrical tortuosity model of Ghanbarian et al. (2013b) (since $\tau_g \leq \tau_e$), one may expect Eq. (9) combined with Eq. (17) to overestimate $\sigma_b/\sigma_f$ in natural rocks. Another source of error might be due to estimating $\langle r_b^2 \rangle$ from the pore-throat size distribution rather than pore-body size distribution. In rocks the pore-body size distribution is not necessarily the same as the pore-throat size distribution. Therefore, replacing $\langle r_b^2 \rangle$ with $\langle r^2 \rangle$ may yield uncertainties in the estimation of electrical conductivity and permeability. Accurate $\sigma_b/\sigma_f$ estimate also depends on the value of $r_{ee}$ whose overestimation in turn leads to overestimation of $\sigma_b/\sigma_f$.

Figure 3b shows the estimated permeability via Eq. (10) with hydraulic tortuosity $\tau_h$ estimated from Eq. (17) as a function of the measured one. As can be observed, although for most samples Eq. (10) estimated *k* from the estimated tortuosity and the pore-throat size distribution derived from the MICP curve reasonably well, it underestimated *k* in some samples. The general trend of *k* underestimation via Eq. (10) is confirmed through the negative MLE value (i.e., -0.36 reported in Fig. 3b). Given that Eq. (17) probably underestimated $\tau_h$, *k* underestimations via Eq. (10) might be because of either inaccurate values of $\langle r_b^2 \rangle$ approximated by $\langle r^2 \rangle$ or imprecise pore shape factor 8.

In Fig. 3c we presented the calculated gas permeability *k* via Eq. (12) from the pore-throat size distribution and measured electrical conductivity data versus the measured *k*. As shown in Fig. 3c, Eq. (12) tended to overestimate *k* slightly. However, most estimations are within a factor of two confidence intervals. The calculated RMSLE and MLE values given in Fig. 3 indicate that Eq. (12) estimated *k* more accurately than Eq. (10) that estimates *k* from the pore-throat size distribution and tortuosity. This is mainly because the value of tortuosity was approximately estimated, while the electrical conductivity value was experimentally measured.

In Fig. 4 we show results for the same tight-gas sandstone samples but assuming pores are slit-shaped. Similar to the results of cylindrical pores (Fig. 3a) as Fig. 4a shows the EMA, Eq. (13), overestimated $\sigma_b/\sigma_f$ (MLE = 0.61). However, comparing the calculated RMSLE values (0.62 vs. 0.53) clearly shows that Eq. (9) estimated $\sigma_b/\sigma_f$ more precisely than Eq. (13). One should note that all parameters in Eqs. (9) and (13) are the same except for the effective electrical pore size (i.e., $r_{ee}$ and $w_{ee}$) and their exponents. As can be deduced from Table 1, the value of $w_{ee}$ is less than that of $r_{ee}$ for all samples ($w_{ee} < r_{ee}$). More specifically, we found $w_{ee} = 0.947 r_{ee} + 0.003$ with $R^2 = 0.99$ and, accordingly, $w_{ee}$ in Eq. (13) greater than $r_{ee}^2$ in Eq. (13) ($w_{ee} > r_{ee}^2$). This is the reason that Eq. (13) gives a larger and less accurate estimate of $\sigma_b/\sigma_f$ than Eq. (9).

Figure 4b presents the permeability calculated via Eq. (14) versus the measured one. As can be observed, most $k$ estimations are near one order of magnitude greater than the measurements (MLE = 0.55). Comparing Fig. 4b with Fig. 3b and the calculated RMSLE values (0.74 vs. 0.61) indicate that Eq. (10) substantially estimated $k$ more accurately than Eq. (14). These two models differ in two parameters: (1) the pore shape factor i.e., 8 and 12, and (2) and the effective pore size i.e., $r_{he}^4$ and $w_{he}^3$ for cylindrical and slit-shaped pores, respectively. We experimentally found $w_{he} = 0.967 r_{he} + 0.003$ with $R^2 = 0.99$ ($w_{he} < r_{he}$), which means $w_{he}^3 > r_{he}^4$. Although the pore shape factor 8 in Eq. (10) is less than 12 in Eq. (14) by a factor of 1.5, because $w_{he}^3$ is greater than $r_{he}^4$ by a factor between 2 and 16, $k$ estimates via Eq. (14) are greater than those via Eq. (10). As we discussed earlier, the high correlation between $w_{he}$ and $r_{he}$ is mainly because both are estimated from Eq. (8) with the same input parameters but the conductance-size exponents are different for the hydraulic (i.e., 4) and electrical (i.e., 3) flow.

Results of electrical conductivity and permeability estimates from the pore-throat size distribution, tortuosity, and porosity (i.e., Eqs. (9), (10), (13) and (14)) demonstrate that cylindrical pores might be more realistic than slit-shaped pores and yield more accurate estimations in tight-gas sandstones. However, in the following we show that the EMA (e.g., Eqs. (12) and (15)) with slit-shaped pores provides slightly more precise permeability estimations from the pore-throat size distribution and measured electrical conductivity than cylindrical pores.

In Fig. 4c we show the calculated permeability using Eq. (15) as a function of the measured $k$. For all samples except one, the estimated permeabilities are within a factor of two confidence intervals, which indicates that the EMA estimations from the pore-throat size distribution and measured electrical conductivity are in reasonable agreement with the experimentally measured values (MLE = -0.03). Comparing Figs. 4c with 3c shows that Eq. (15), assuming pores are slit-shaped, estimated $k$ with RMSLE = 0.42 slightly more precisely than Eq. (12), with RMSLE = 0.46, in which pores are cylindrical. These results indicate that accurate electrical conductivity and permeability estimations do not necessarily provide useful information and precise conclusions about representative pore geometries in rocks. We should point out that rock minerals play a critical role in the formation of pore space and its structure. Different mineralogical compositions could greatly affect the pore shape geometry. Some rocks include minerals which are not stable under a wide range of pressure and temperature conditions, whereby alteration mechanisms such as cementation, dissolution, and recrystallization (Vanorio et al., 2008) are responsible for the presence of pores with irregular shapes.

Comparison of the RMSLE values reported in Fig. 3a,b to those in Fig. 4a,b confirms that the assumption of cylindrical pores yielded more accurate electrical conductivity and permeability estimates from pore-throat size distribution, porosity and tortuosity than the assumption of slit-shaped pores. However, the EMA with slit-shaped pores estimated permeability from pore-throat size distribution and formation factor more precisely than that with cylindrical pores (compare Fig. 3c to Fig. 4c).

**4.2. Scaling up via critical path analysis**

Another upscaling technique from statistical physics is critical path analysis (CPA). In contrast to the effective-medium approximation that is known to be accurate in porous media with relatively narrow pore-throat size distribution (Adler and Berkowitz, 2000), CPA works well in highly heterogeneous media whose pore-throat size distribution is broad. In what follows, we compare the accuracy of CPA with that of the EMA in the estimation of gas permeability from the pore-throat size distribution and measured electrical conductivity using the same eighteen tight-gas sandstone samples.

Based on concepts from CPA, Le Doussal (Le Doussal, 1989) and later Skaggs (Skaggs, 2011) proposed

$$k = \frac{2^{-y}}{8} \frac{\sigma_b}{\sigma_f} r_c^2 \qquad (19)$$

when pores are cylindrical, and

$$k = \frac{3^{-y}}{12} \frac{\sigma_b}{\sigma_f} w_c^2 \qquad (20)$$

assuming pores are slit-shaped. In Eqs. (19) and (20), $r_c$ and $w_c$ represent critical pore-throat radius and width, respectively, and $y = 0.74$ (Skaggs, 2003). Both $r_c$ and $w_c$ can be approximated from the inflection point on the MICP curve (Katz and Thompson, 1986; Katz and Thompson, 1987).

Results of $k$ estimations via Eqs. (19) and (20) are presented in Fig. 5 assuming pores are cylindrical (Fig. 5a) and slit-shaped (Fig. 5b). As can be deduced from the RMSLE and MLE values reported in Fig. 5, Eq. (19) estimated $k$ more accurately than Eq. (20), in contrast to the EMA results presented above. By comparing Figs. (3c) and (4c) we found that the EMA with slit-shaped pores was more precise than that with cylindrical pores. This contradiction clearly shows that, in addition to input parameters, the choice of upscaling techniques would affect permeability estimation accuracy as well as any conclusion about representative pore geometry in the medium.

We also found that the EMA, either Eq. (12) or Eq. (15) estimated $k$ from the pore-throat size distribution and measured electrical conductivity with slightly more accuracy than CPA, Eqs. (19) and (20) (compare RMSLE = 0.46 and 0.42 with 0.47 and 0.65). Although the reliability of Eq. (19) is not substantially less than that of Eqs. (12) and (15), the results show that in these tight-gas sandstones with relatively narrow pore-throat size distribution (lognormal pore-throat size distribution standard deviation $s \leq 0.9$; see Table 1) the EMA returned more accurate $k$ estimations than CPA.

Based on electrical conductivity simulations of Adler and Berkowitz (2000), one should expect very accurate estimations via the EMA in three dimensions, if the standard deviation of the lognormal electrical conductance distribution in a network of pores is less than 1 ($s_{ge} < 1$). Given that $f(g_e)dg_e = f(r)dr = f(w)dw$, converting the electrical conductance distribution, $f(g_e)$, to the pore-throat size distribution, $f(r)$ or $f(w)$, requires invoking the relationship between pore electrical conductance, $g_e$, and pore-throat size, $r$ or $w$ (see Eqs. 4 and 6). If one assumes that pores are cylindrical and applies Eq. (4), then $s_{ge} \approx 2s < 1$, while assuming slit-shaped pores and using Eq. (6) result in $s_{ge} \approx s < 1$. A natural porous medium, however, is probably a mixture of the two pore geometries. Accordingly, based on the Adler and Berkowitz (2000) results (i.e., $s_{ge}$ should be less than 1), one may use an average value and expect accurate estimates by the EMA in porous media whose $s$ values are less than 0.75. Interestingly, for all samples analyzed in this study $s \lesssim 0.75$ (see Table 1), except sample 10 for which $s$ is not noticeably greater than 0.75. We should point out that Shah and Yortsos (1996) also argued that porous media, even though moderately disordered in pore size, possess a wider electrical or hydraulic conductance distribution compared to their pore-throat size distribution.

## 5. Conclusions

Using the effective-medium approximation we scaled up electrical conductivity, $\sigma_b/\sigma_f$, and gas permeability, $k$, from pore to core in tight-gas sandstones. More specifically, we evaluated various effective-medium approximation models in the estimation of $\sigma_b/\sigma_f$ and $k$ by assuming different pore geometries e.g., cylindrical and slit-shaped. Comparison of eighteen tight-gas sandstone samples indicated that the EMA estimated $\sigma_b/\sigma_f$ and $k$ from pore-throat size distribution, porosity and tortuosity more accurately by assuming cylindrical pores. We also compared the accuracy of the effective-medium approximation to the reliability of critical path analysis in the estimation of $k$ from the pore-throat size distribution and the measured electrical conductivity. Results showed that the effective-medium approximation with slit-shaped pores estimated $k$ (RMSLE = 0.42) slightly more precisely than critical path analysis with cylindrical pores (RMSLE = 0.47), although both method estimations were mainly within a factor of two of the measurements. The reason most probably is because the width of the pore-throat size distribution of these tight-gas sandstones is relatively narrow, consistent with concepts of the EMA. We found that, depending on input parameters and upscaling methods, the assumption of cylindrical pores could yield more accurate $\sigma_b/\sigma_f$ and $k$ estimates than the assumption of slit-shaped pores and vice versa. This means that accurate estimates of electrical conductivity and permeability do not provide a reliable way to distinguish between models of the representative pore geometry.

**Acknowledgements**

BG is grateful to Kansas State University for supports through faculty startup funds. Larry W. Lake holds the Shahid and Sharon Ullah Chair and Carlos Torres-Verdín holds the Brian James Jennings Memorial Endowed Chair, both at the University of Texas at Austin.

**Notation**

| | |
|---|---|
| $b$ | slit-shaped pore breadth [μm] |
| $C$ | typical pore-throat length [μm] |
| $d$ | pore-throat diameter [μm] |
| $\bar{d}_g$ | representative grain diameter [m] |
| $D_{opt}$ | optimal path fractal dimension [-] |
| $f(g)$ | conductance distribution [-] |
| $f(g_e)$ | electrical conductance distribution [-] |
| $f(r)$ | cylindrical pore-throat size distribution [-] |
| $f(w)$ | slit-shaped pore-throat size distribution [-] |
| $F$ | formation factor [-] |
| $g$ | pore conductance |
| $g_e$ | pore electrical conductance [S m$^{-1}$ μm] |
| $g_h$ | pore hydraulic conductance [μm$^3$ Pa$^{-1}$ s$^{-1}$] |
| $g_{min}$ | minimum pore conductance |
| $g_{max}$ | maximum pore conductance |
| $k$ | Klinkenberg-corrected gas permeability [μm$^2$] |
| $l$ | pore length [μm] |
| $L_e$ | effective geometrical flow length [μm] |
| $L_s$ | sample length [μm] |
| $L_t$ | tortuous capillary tube length [μm] |
| $p_c$ | percolation threshold [-] |
| $P_c$ | capillary pressure [kPa] |
| $P_{nw}$ | nonwetting phase pressure [kPa] |
| $P_w$ | wetting phase pressure [kPa] |
| $r$ | pore-throat radius [μm] |
| $r_b$ | pore-body radius [μm] |
| $\langle r_b^2 \rangle$ | average squared pore-body radius [μm$^2$] |
| $r_{50}$ | median pore-throat radius [μm] |
| $r_c$ | critical cylindrical pore-throat radius [μm] |
| $r_e$ | effective pore-throat radius [μm] |
| $r_{ee}$ | effective electrical pore-throat radius [μm] |
| $r_{he}$ | effective hydraulic pore-throat radius [μm] |
| $R$ | capillary tube radius [μm] |
| $\bar{R}$ | representative pore radius in the medium [μm] |
| $R_h$ | hydraulic radius [μm] |
| $s$ | standard deviation of lognormal pore-throat size distribution [-] |
| $s_{ge}$ | standard deviation of lognormal electrical conductance distribution [-] |
| $S_a$ | surface area [m$^{-1}$] |
| $S_c$ | critical saturation [-] |
| $S_{Hg}$ | mercury saturation [-] |
| $w$ | slit-shaped pore-throat width [μm] |
| $w_b$ | slit-shaped pore-body width [μm] |
| $w_c$ | critical slit-shaped pore-throat width [μm] |
| $w_{ee}$ | effective electrical slit-shaped pore-throat width [μm] |
| $w_{he}$ | effective hydraulic slit-shaped pore-throat width [μm] |
| $y$ | scaling exponent from critical path analysis [-] |
| $Z$ | average pore coordination number [-] |
| | |
| $\phi$ | porosity [cm$^3$ cm$^{-3}$] |
| $\gamma$ | air/mercury interfacial tension [dyn cm$^{-1}$] |
| $\mu$ | fluid viscosity [Pa s] |
| $\sigma_b$ | bulk electrical conductivity [S m$^{-1}$] |
| $\sigma_f$ | fluid electrical conductivity [S m$^{-1}$] |
| $\tau$ | tortuosity [-] |
| $\tau_e$ | electrical tortuosity [-] |
| $\tau_g$ | geometrical tortuosity [-] |
| $\tau_h$ | hydraulic tortuosity [-] |
| $\theta$ | contact angle [°] |
| $\nu$ | correlation length exponent [-] |

**Figure captions**

Figure 1. Three-dimensional schematic of a disordered pore network (left) replaced by an ordered network (right) with an effective pore size (a modified version of Doyen (1988)).

Figure 2. Pore-throat size distribution, determined from the measured mercury intrusion capillary pressure (MICP), the interpolated curve using the spline method, and the effective hydraulic and electrical pore widths for samples 3, 8, 13 and 17. $S_{Hg}$ and $P_c$ respectively represent mercury saturation and capillary pressure.

Figure 3. Logarithm of the calculated electrical conductivity and gas permeability assuming that pores are cylindrical and using various models from the effective-medium approximation vs. the measured one for eighteen tight-gas sandstones. Electrical conductivity and permeability were estimated from the pore-throat size distribution and estimated geometrical tortuosity in plots (a) and (b) and from the pore-throat size distribution and measured electrical conductivity in plot (c). The red solid and dashed red lines represent the 1:1 line ($y = x$) and factor of two confidence intervals ($y = 0.5x$ and $2x$), respectively.

Figure 4. Logarithm of the calculated electrical conductivity and gas permeability assuming that pores are slit-shaped and using various models from the effective-medium approximation vs. the measured one for eighteen tight-gas sandstones. Electrical conductivity and permeability were estimated from the pore-throat size distribution and estimated geometrical tortuosity in plots (a) and (b) and from the pore-throat size distribution and measured electrical conductivity in plot (c). The red solid

and dashed red lines represent the 1:1 line ($y = x$) and factor of two confidence intervals ($y = 0.5x$ and $2x$), respectively.

Figure 5. Logarithm of the calculated gas permeability using critical path analysis (CPA) and assuming that pores are (a) cylindrical and (b) slit-shaped, respectively, vs. the measured one for eighteen tight-gas sandstones. Permeability was estimated from the pore-throat size distribution and measured electrical conductivity. The red solid and dashed red lines represent the 1:1 line ($y = x$) and factor of two confidence intervals ($y = 0.5x$ and $2x$), respectively.

Table 1. Salient properties of tight-gas sandstone samples used in this study.

| Sample | $\phi^*$ | $\sigma_b/\sigma_f$ ×10³ | $k$ [μm²] ×10⁶ | Lognormal distribution $\mu$ | Lognormal distribution $s$ | $S_c$ | $r_c$ [μm] | $r_{ee}$ [μm] | $r_{he}$ [μm] | $w_{ee}$ [μm] | $w_{he}$ [μm] | $\langle r_b \rangle$ [μm] | $\langle r_b^2 \rangle$ [μm²] | $\tau_g$ Eq. (17) |
|---|---|---|---|---|---|---|---|---|---|---|---|---|---|---|
| 1 | 0.068 | 2.91 | 10.86 | -1.49 | 0.22 | 0.43 | 0.194 | 0.216 | 0.221 | 0.212 | 0.219 | 0.125 | 0.041 | 11.5 |
| 2 | 0.074 | 3.13 | 13.82 | -1.41 | 0.19 | 0.30 | 0.235 | 0.237 | 0.244 | 0.230 | 0.241 | 0.116 | 0.035 | 9.3 |
| 3 | 0.086 | 3.87 | 29.61 | -1.18 | 0.27 | 0.27 | 0.298 | 0.321 | 0.333 | 0.308 | 0.329 | 0.219 | 0.147 | 8.0 |
| 4 | 0.072 | 3.93 | 24.67 | -1.65 | 0.37 | 0.41 | 0.159 | 0.213 | 0.211 | 0.211 | 0.212 | 0.209 | 0.129 | 10.7 |
| 5 | 0.089 | 4.46 | 39.48 | -1.00 | 0.16 | 0.06 | 0.366 | 0.398 | 0.446 | 0.368 | 0.424 | 0.247 | 0.181 | 6.5 |
| 6 | 0.079 | 3.68 | 17.76 | -1.22 | 0.21 | 0.31 | 0.279 | 0.280 | 0.291 | 0.272 | 0.287 | 0.134 | 0.048 | 8.9 |
| 7 | 0.069 | 3.27 | 17.76 | -1.22 | 0.16 | 0.26 | 0.284 | 0.275 | 0.287 | 0.266 | 0.282 | 0.125 | 0.041 | 9.4 |
| 8 | 0.077 | 3.44 | 20.72 | -1.10 | 0.16 | 0.32 | 0.305 | 0.307 | 0.320 | 0.296 | 0.315 | 0.157 | 0.062 | 9.2 |
| 9 | 0.067 | 2.49 | 6.91 | -1.64 | 0.58 | 0.22 | 0.176 | 0.174 | 0.186 | 0.164 | 0.181 | 0.085 | 0.018 | 9.3 |
| 10 | 0.057 | 3.02 | 2.96 | -1.66 | 0.90 | 0.28 | 0.123 | 0.131 | 0.141 | 0.123 | 0.137 | 0.052 | 0.007 | 11.1 |
| 11 | 0.083 | 4.32 | 12.83 | -1.32 | 0.33 | 0.25 | 0.254 | 0.245 | 0.259 | 0.234 | 0.253 | 0.107 | 0.032 | 8.0 |
| 12 | 0.084 | 3.91 | 8.88 | -1.60 | 0.29 | 0.25 | 0.196 | 0.192 | 0.202 | 0.185 | 0.198 | 0.098 | 0.025 | 8.0 |
| 13 | 0.043 | 1.24 | 1.18 | -2.07 | 0.56 | 0.36 | 0.102 | 0.117 | 0.123 | 0.113 | 0.121 | 0.06 | 0.008 | 15.0 |
| 14 | 0.073 | 3.41 | 4.93 | -1.87 | 0.60 | 0.24 | 0.140 | 0.136 | 0.145 | 0.129 | 0.141 | 0.072 | 0.013 | 8.8 |
| 15 | 0.073 | 4.12 | 25.66 | -1.57 | 0.44 | 0.28 | 0.211 | 0.210 | 0.220 | 0.201 | 0.216 | 0.130 | 0.047 | 9.1 |
| 16 | 0.050 | 2.05 | 38.49 | -2.56 | 0.56 | 0.30 | 0.066 | 0.069 | 0.074 | 0.066 | 0.072 | 0.033 | 0.002 | 12.5 |
| 17 | 0.062 | 3.39 | 8.88 | -1.91 | 0.58 | 0.21 | 0.148 | 0.134 | 0.144 | 0.127 | 0.140 | 0.067 | 0.011 | 9.7 |
| 18 | 0.069 | 2.99 | 1.97 | -2.85 | 0.52 | 0.41 | 0.046 | 0.060 | 0.061 | 0.058 | 0.061 | 0.041 | 0.004 | 11.1 |

*$\phi$: porosity, $\sigma_b/\sigma_f$: electrical conductivity, $k$: *Klinkenberg-corrected* permeability, $\mu$: lognormal distribution mean, $s$: lognormal pore-throat size distribution standard deviation, $r_c = w_c$: critical pore-throat radius/width, $r_{ee}$: effective electrical pore-throat radius, $r_{he}$: effective hydraulic pore-throat radius, $w_{ee}$: effective electrical pore-throat width, $w_{he}$: effective hydraulic pore-throat width, $\langle r_b \rangle = \langle w_b \rangle$: spatial average of pore body, $\langle r_b^2 \rangle = \langle w_b^2 \rangle$: spatial average of squared pore body, $\tau_g = (L_e/L_s)^2$: geometrical tortuosity.

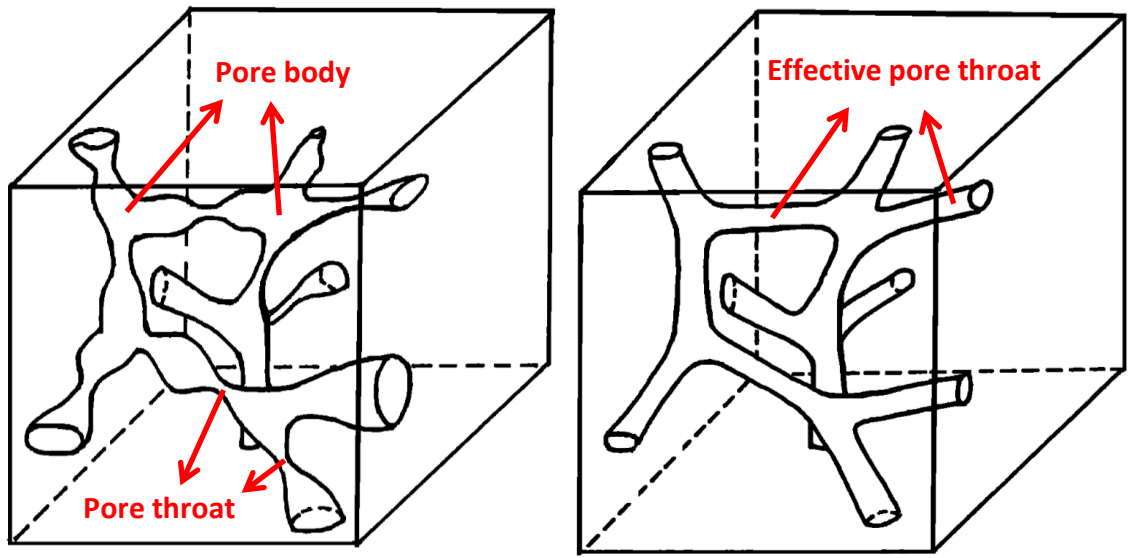

Figure 1. Three-dimensional schematic of a disordered pore network (left) replaced by an ordered network (right) with an effective pore size (a modified version of Doyen (1988)).

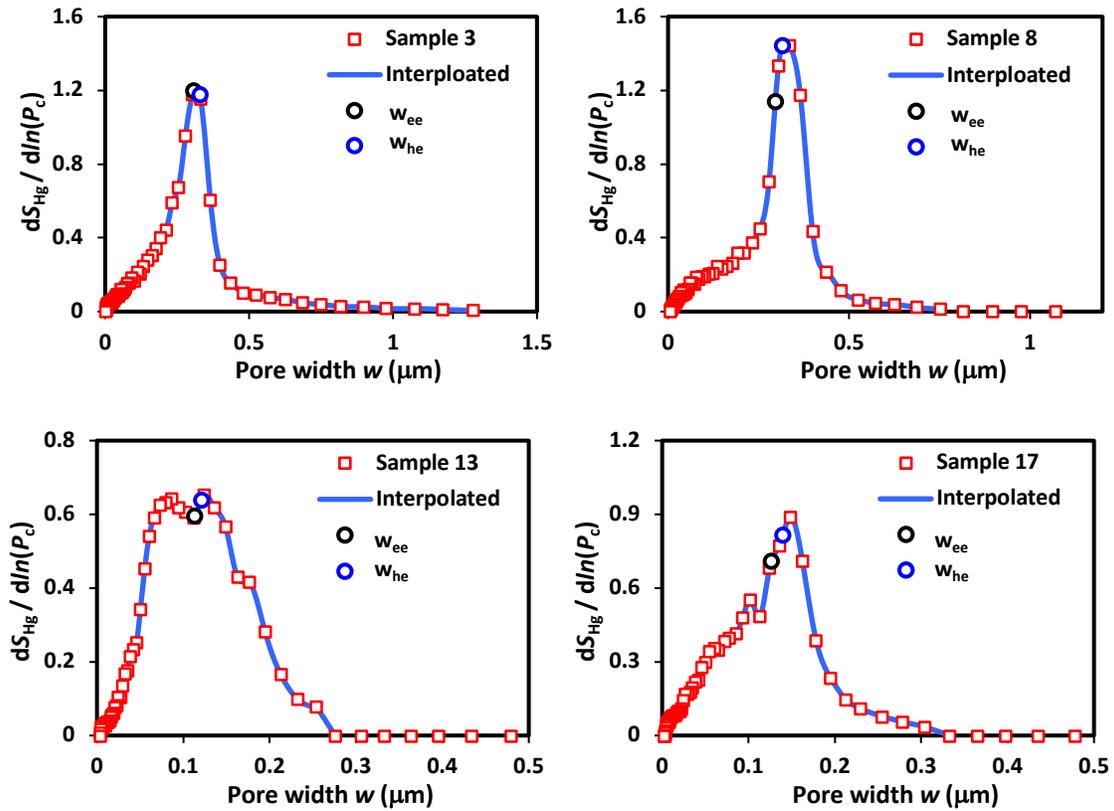

Figure 2. Pore-throat size distribution, determined from the measured mercury intrusion capillary pressure (MICP), the interpolated curve using the spline method, and the effective hydraulic and electrical pore widths for samples 3, 8, 13 and 17. $S_{Hg}$ and $P_c$ respectively represent mercury saturation and capillary pressure.

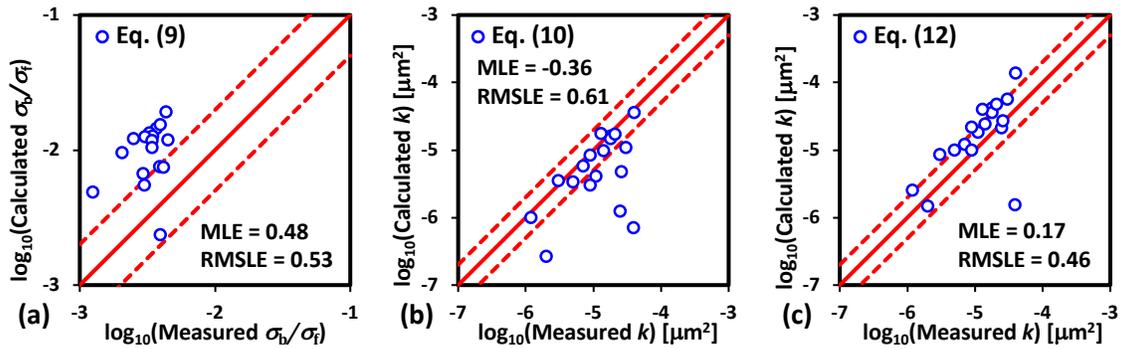

Figure 3. Logarithm of the calculated electrical conductivity and gas permeability assuming that pores are *cylindrical* and using various models from the *effective-medium approximation* vs. the measured one for eighteen tight-gas sandstones. Electrical conductivity and permeability were estimated from the pore-throat size distribution and estimated geometrical tortuosity in plots (a) and (b) and from the pore-throat size distribution and measured electrical conductivity in plot (c). The red solid and dashed red lines represent the 1:1 line ($y = x$) and factor of two confidence intervals ($y = 0.5x$ and $2x$), respectively.

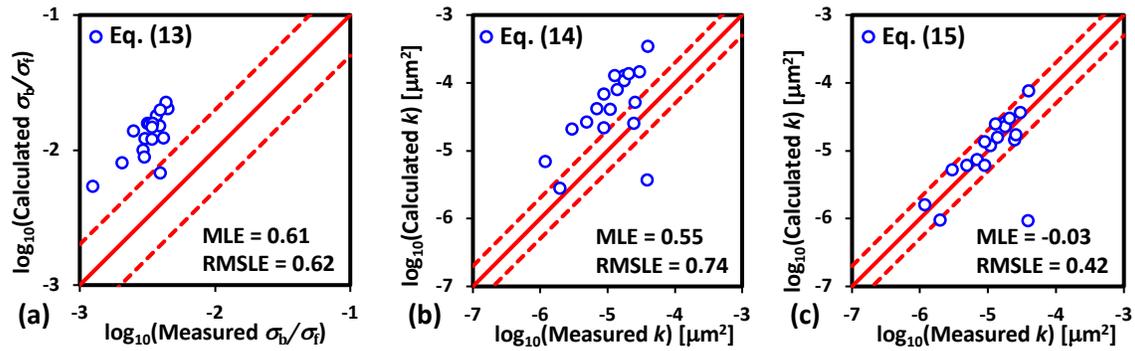

Figure 4. Logarithm of the calculated electrical conductivity and gas permeability assuming that pores are *slit-shaped* and using various models from the *effective-medium approximation* vs. the measured one for eighteen tight-gas sandstones. Electrical conductivity and permeability were estimated from the pore-throat size distribution and estimated geometrical tortuosity in plots (a) and (b) and from the pore-throat size distribution and measured electrical conductivity in plot (c). The red solid and dashed red lines represent the 1:1 line ($y = x$) and factor of two confidence intervals ($y = 0.5x$ and $2x$), respectively.

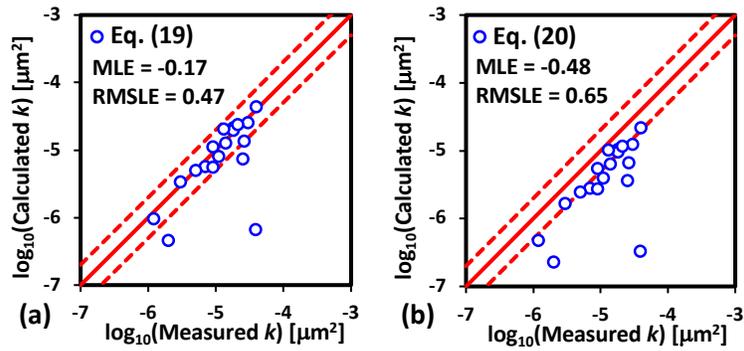

Figure 5. Logarithm of the calculated gas permeability using critical path analysis (CPA) and assuming that pores are (a) cylindrical and (b) slit-shaped, respectively, vs. the measured one for eighteen tight-gas sandstones. Permeability was estimated from the pore-throat size distribution and measured electrical conductivity. The red solid and dashed red lines represent the 1:1 line ($y = x$) and factor of two confidence intervals ($y = 0.5x$ and $2x$), respectively.